\documentclass[conference]{IEEEtran}
\IEEEoverridecommandlockouts
\usepackage{cite}
\usepackage{amsmath,amssymb,amsfonts}
\usepackage{algorithmic}
\usepackage{graphicx}
\usepackage{textcomp}
\usepackage{url}
\usepackage{lipsum}
\usepackage{multirow}
\usepackage{bm}
\usepackage{makecell}
\def\glnref#1{Eq.~(\eqref{#1})}
\def\figref#1{Fig.~\ref{#1}}
\def\tabref#1{Table~\ref{#1}}

\usepackage{xcolor}
\def\BibTeX{{\rm B\kern-.05em{\sc i\kern-.025em b}\kern-.08em
    T\kern-.1667em\lower.7ex\hbox{E}\kern-.125emX}}
\begin{document}

\title{Understanding Hyperspherical Geometry of ECAPA-TDNN Embedding \\ and Its Impact on Zero-Shot Voice Conversion
}

\author{\IEEEauthorblockN{Mathilde Abrassart}
\IEEEauthorblockA{\textit{STMS Lab, IRCAM} \\
\textit{CNRS, Sorbonne Université}\\
Paris, France \\
mathilde.abrassart@ircam.fr}
\and
\IEEEauthorblockN{Nicolas Obin}
\IEEEauthorblockA{\textit{STMS Lab, IRCAM} \\
\textit{CNRS, Sorbonne Université}\\
Paris, France \\
nicolas.obin@ircam.fr}
\and
\IEEEauthorblockN{Axel Roebel}
\IEEEauthorblockA{\textit{STMS Lab, IRCAM} \\
\textit{CNRS, Sorbonne Université}\\
Paris, France \\
axel.roebel@ircam.fr}}

\maketitle

\begin{abstract}
Angular-margin speaker encoders are widely used in voice conversion, yet the geometry of their classifier prototypes remains poorly understood. We analyze ECAPA-TDNN classifier prototypes as points on the unit hypersphere and characterize their organization using rotation-invariant angular statistics together with global and local effective dimensionality measures. Our analysis shows that standard training can induce angular concentration and a substantial reduction in effective dimensionality. To address this, we investigate two geometric regularization strategies (hinged Riesz log-energy and effective-dimension maximization) applied to classifier prototypes to encourage more uniform hyperspherical coverage. The resulting prototype sets exhibit higher effective dimensionality and improved isotropy, with configuration-dependent effects on speaker-recognition performance. When the corresponding ECAPA-TDNN models are used as speaker encoders for Fast-VGAN, the regularized systems also exhibit improved robustness in zero-shot voice conversion, particularly for previously unseen speakers.

\end{abstract}


\section{Introduction}
Speaker embeddings have become a fundamental component of modern speech processing systems, enabling robust speaker verification, diarization, and increasingly conditional generative modeling. Early approaches such as i-vectors \cite{dehak2010front} relied on generative factor analysis, while the introduction of discriminative neural embeddings such as x-vectors \cite{snyder2018x} marked a shift toward supervised representation learning. Subsequent architectural refinements, including ECAPA-TDNN \cite{desplanques2020ecapa}, further improved discriminative power on large-scale benchmarks. Beyond speaker recognition, these embeddings are widely used as conditioning signals for multi-speaker speech synthesis and zero-shot voice conversion, where generalization to unseen speakers is critical.

In parallel, training objectives evolved from standard cross-entropy classification toward angular-margin-based losses rooted in hyperspherical metric learning, such as ArcFace \cite{deng2019arcface}. These objectives normalize embeddings and classifier prototypes onto the unit hypersphere, making classification depend solely on angular separation. Both open-set speaker verification and zero-shot voice conversion require generalization to unseen speakers. However, zero-shot voice conversion additionally relies on the speaker representation as a conditioning space, for which smooth and well-covered regions may facilitate the representation of speakers that were not observed during training. We therefore investigate whether improving the isotropy and coverage of this space benefits zero-shot voice conversion.

Recent advances in representation learning highlight the importance of geometric organization in such settings. The alignment-uniformity framework \cite{wang2020understanding} shows that effective embeddings balance intra-class compactness and hyperspherical coverage, while Neural Collapse predicts convergence of class means toward simplex equiangular configurations in late-stage training \cite{papyan2020prevalence}. Several studies have also related intrinsic dimensionality and spectral properties to representation quality \cite{ansuini2019intrinsic,recanatesi2019dimensionality,martin2021implicit}. Despite the widespread use of angular-margin objectives in speaker embedding systems, relatively little attention has been devoted to understanding the geometry of the resulting classifier prototypes or to evaluating how this geometry affects downstream applications such as zero-shot voice conversion.

\noindent This paper makes the following contributions:

\begin{enumerate}
    \item We present a systematic geometric analysis of the hyperspherical prototype spaces induced by angular-margin speaker encoders or as conditioning representations in generative models, characterizing their structure through global and local effective dimensionality and angular distance statistics.

    \item Motivated by the observed angular concentration and reduced effective dimensionality, we introduce two geometric regularizers that encourage a more uniform hyperspherical distribution and higher effective dimensionality of classifier prototypes.

    \item We provide a preliminary evaluation of the practical implications of these geometric properties in two settings: speaker classification and zero-shot voice conversion using Fast-VGAN \cite{abrassart2025fast}, where ECAPA embeddings are used as conditioning representations for the generative model. The regularized systems exhibit improved zero-shot voice conversion robustness together with substantially more isotropic and higher-dimensional prototype spaces, with configuration-dependent speaker-recognition results.
\end{enumerate}

\section{Related Work}

\subsection{Speaker representation learning with angular margins}

Speaker verification has progressively shifted from probabilistic latent-variable modeling to fully discriminative representation learning. The introduction of x-vectors \cite{snyder2018x} marked a transition to supervised neural embeddings, where a TDNN with statistical pooling learns utterance-level speaker representations.
Recent systems incorporate stronger geometric constraints through $\ell_2$ normalization and angular-margin objectives, projecting both embeddings $\mathbf{x}$ and classifier weights $\mathbf{w}_k$ onto the unit hypersphere,
$\lVert \mathbf{x}\rVert_2 = \lVert\mathbf{w}_k\rVert_2 = 1$,
so that classification depends only on the cosine similarity
$\cos\theta_k = \mathbf{w}_k^\top \mathbf{x}$.
This paradigm was first explored in face recognition through SphereFace \cite{liu2017sphereface}, CosFace \cite{wang2018cosface}, and ArcFace \cite{deng2019arcface}, and was subsequently adopted in speaker verification \cite{li2018angular, yu2019ensemble}. Modern ECAPA-TDNN systems commonly employ ArcFace-like objectives \cite{desplanques2020ecapa, thienpondt2020idlab}.
ECAPA-TDNN further improved speaker embeddings through multi-scale temporal modeling, channel attention, and attentive pooling, achieving strong performance on VoxCeleb \cite{desplanques2020ecapa}. Numerous extensions have since improved robustness and discriminability \cite{yao2023branch, liu2023ecapa++, chen2022large}, and ECAPA embeddings have been successfully transferred to tasks such as diarization and multi-speaker TTS \cite{dawalatabad2021ecapa, xue2022ecapa}.
Despite strong downstream performance (e.g., EER, DER), the geometric structure of the embedding and classifier spaces—its isotropy, dimensionality, and angular organization—remains largely unexplored, motivating our study.

\subsection{Geometric analysis of representation spaces}

The geometry of learned representations has become a central topic in deep learning. In contrastive learning, the alignment-uniformity framework shows that effective embeddings balance intra-class compactness with uniform hyperspherical coverage \cite{wang2020understanding}. In supervised settings, Neural Collapse predicts convergence of class means toward a simplex equiangular configuration during late training \cite{papyan2020prevalence}. Several studies further analyze the intrinsic dimensionality of deep representations, reporting substantial dimensional compression and structured anisotropy across layers \cite{ansuini2019intrinsic,recanatesi2019dimensionality}. Complementary spectral analyses relate eigenvalue structure to implicit regularization and effective rank in deep networks \cite{martin2021implicit}. More recently, hyperspherical energy objectives have been proposed to promote uniformly distributed embeddings on the unit sphere \cite{liu2018learning}, while participation-ratio-based measures have also been investigated both as representation quality indicators and optimization objectives \cite{kuznetsov2026information}. While these geometric phenomena are well established in vision and theoretical deep learning, they remain largely unexplored in the speech domain, particularly for speaker verification and zero-shot voice conversion. We address this gap by explicitly characterizing the isotropy, intrinsic dimensionality, and angular organization of ECAPA classifier prototypes trained with angular margins, and by studying how geometry-aware regularization affects downstream performance.

\section{Geometry of the ECAPA Prototype Space}
\label{sec:geom}

In this section, we analyze the geometry of the ECAPA-TDNN speaker prototypes obtained by training an ECAPA-TDNN model on LibriTTS-R \cite{koizumi2023libritts} derived from LibriTTS \cite{zen2019libritts}. Our goal is therefore not to maximize isotropy for its own sake, but to test whether reducing prototype-space anisotropy improves hyperspherical coverage and ultimately benefits downstream zero-shot generation. We use the small ECAPA-TDNN variant introduced in \cite{desplanques2020ecapa}. Since our goal is not to achieve robust speaker identification, but to study the geometry of the resulting speaker prototypes and their impact on zero-shot voice conversion, we use Mel-spectrograms as input features, and do not use any of the data augmentation strategies proposed \cite{desplanques2020ecapa}. This choice aligns the speaker encoder with the representation used by the generative model. 

Under the angular-margin classification objective, the columns of the speaker prototype matrix $W \in \mathbb{R}^{D \times C}$ are $\ell_2$-normalized and therefore lie on the unit hypersphere. Classifier prototypes and utterance-level embeddings are distinct and need not share the same distribution. However, the angular-margin objective directly couples them through their cosine similarities. We therefore analyze the prototype geometry explicitly constrained by the classifier, while downstream VC experiments evaluate the representations produced by the corresponding encoder. We characterize the global and local effective dimensionality of the prototype distribution using the participation ratio $D_{PR}$ \cite{recanatesi2022effdim}
\begin{eqnarray}
\Sigma &=& WW^T \\
D_{PR} &=& \frac{Tr(\Sigma)^2}{Tr(\Sigma^2)}
       = \frac{\left(\sum_i \lambda_i\right)^2}
              {\sum_i \lambda_i^2},
\label{eq:eff_dim}
\end{eqnarray}
where $\Sigma$ is the prototype scatter matrix and $\lambda_i$ its eigenvalues, and $Tr(\cdot)$ the trace operator. The trace formulation avoids explicit eigen-decomposition and is therefore efficient to compute as a batch-wise regularizer.
Note that $D_{PR}$ is computed globally ($G_{PR}$) to characterize the dimensionality of the full prototype set and locally ($L_{PR}$), where it is computed over the $k$ nearest neighbors of each prototype. Because the number of speaker prototypes is limited, we use $k=50$, which bounds the locally observable dimensionality by 50.

\subsection{Regularization training strategy}

\begin{figure}[h]
\centering
\includegraphics[width=\columnwidth]{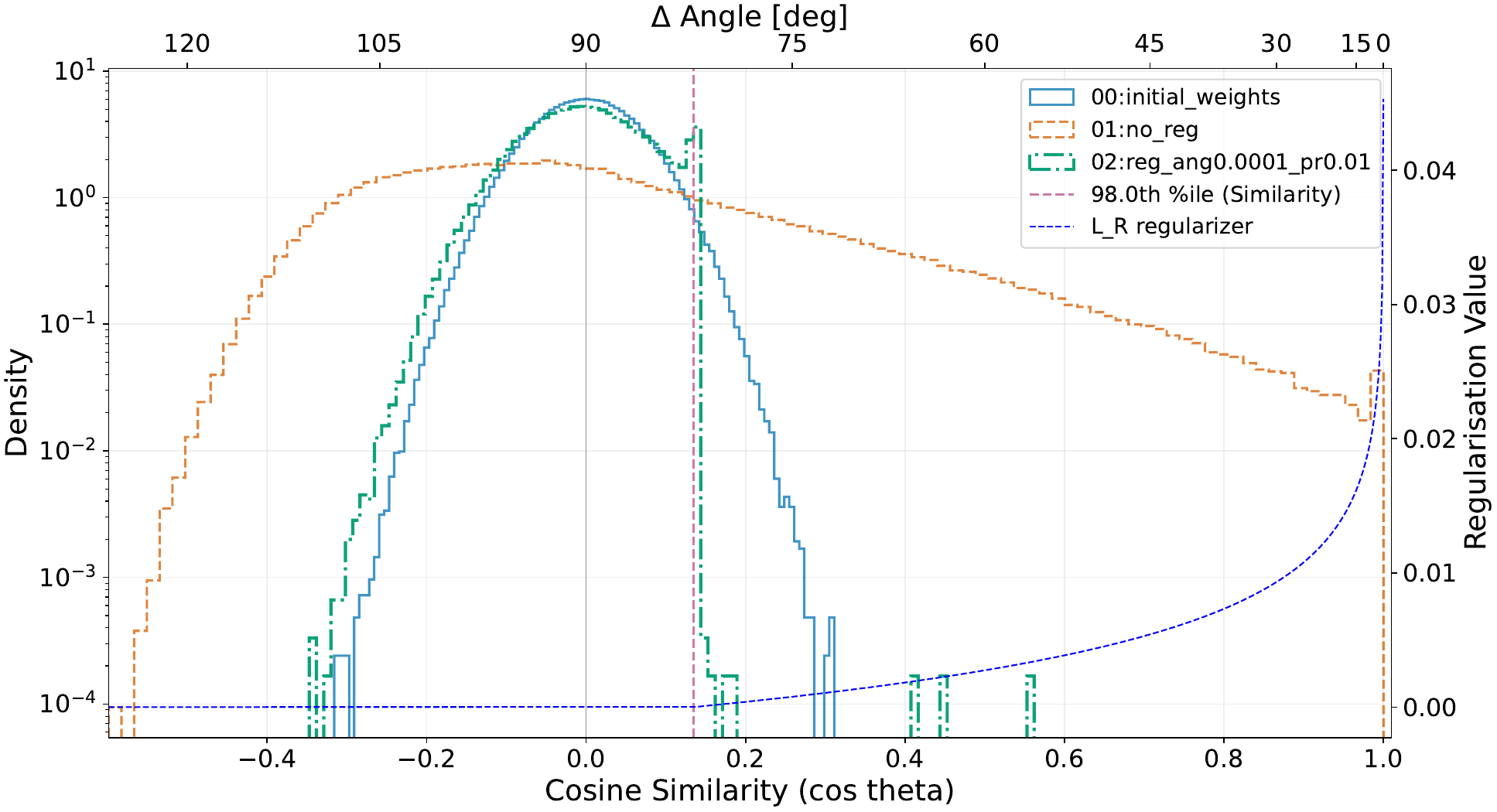}
\vspace{-0.5cm}
\caption{Histogram of pairwise angles between ECAPA-TDNN speaker prototypes directly after initialization, and after training the model on LibriTTS-R with and without regularisation. The regularisation loss is shown in dashed blue.}
\label{fig:reg}
\end{figure}

Although the ArcFace loss promotes inter-speaker separability, its influence on the global organization of the classifier prototypes is mediated by the speakers represented in each training batch. We evaluated batch sizes of 32 and 160 with uniform speaker sampling, corresponding to 32 and 160 distinct speakers per batch. Increasing the batch size had only a limited effect on the resulting prototype geometry, as discussed in Section~\ref{sec:global}. We therefore investigate geometry-aware regularization objectives that act directly on the classifier prototypes, independently of the composition of individual training batches.
The first promotes local angular separation between neighboring prototypes, while the second maximizes the effective dimensionality of the prototype distribution. Together, they improve the global coverage of the hypersphere for zero-shot speaker conditioning. The first regularizer is based on the Riesz log-energy potential \cite{hardin_discretizing_2004,liu2018learning} and aims to prevent excessive angular concentration between neighboring prototypes. We employ the Riesz log energy according to \cite{hardin_discretizing_2004}
\begin{equation}
    R(\omega_N) = -\frac{\sum_{i\ne j} \log(1 - x_i\cdot x_j)}{2N(N-1)}.
     \label{eq:riesz_log_energy}
\end{equation}
Here $\omega_N = \{x_i\}_1^N$, with $\lVert x\rVert = 1$, $x_i\cdot x_j$ denotes cosine similarity, and $N$ is the number of prototypes. Given \eqref{eq:riesz_log_energy} is a global potential and is calculated for all pairs of points with only very few approaching critical angular distances, the direct application of the regularization is rather costly and dilutes the effect of the regularizer due to the small impact of an individual pair. Instead, we use a hinged variant that penalizes only pairs exceeding a cosine similarity threshold $c_t$. Accordingly, this regularizer globally ensures local separability. The threshold depends on the number of points $N$ and the representation dimension $D$. We determine $c_t$ from the distribution of angular distances after weight initialization and use the 98th percentile of cosine similarities among nearest neighbors. The threshold used for our configuration is shown in Figure \ref{fig:reg}. Let $\mathcal{A}$ denote the set of prototype pairs whose cosine similarity exceeds the threshold $c_t$ then the regularizer becomes
\begin{align}
\mathcal{A} &=\{(i,j)\mid i\neq j,\, x_i\cdot x_j>c_t\}.
\label{eq:active_pairs} \\
\mathcal{L}_{\mathrm{R}}
&=
\frac{1}{|\mathcal{A}|}
\sum_{(i,j)\in\mathcal{A}}
-\frac{1}{2}\log
\frac{1-x_i\cdot x_j}{1-c_t}.
\label{eq:hinged_log_energy}
\end{align}

\figref{fig:reg} shows the threshold $c_t$ and the rapid increase of the regularizer loss once the similarity exceeds the threshold.

The second regularizer maximizes the effective dimensionality of the prototype distribution through the participation ratio. To exert sufficient pressure when the effective dimensionality decreases, we define
\begin{equation}
L_{PR} = -\log\left(\frac{D_{PR}}{D} + \epsilon\right).
\end{equation}
Here, $D_{PR}$ denotes the effective dimension from \glnref{eq:eff_dim} and $D$ the prototype dimension.
Full regularization loss becomes 
\begin{equation}
L_{reg} = \lambda_{PR} L_{PR} + \lambda_R L_{R}
\end{equation}

Training the same ECAPA-TDNN model as before, using $\lambda_{PR} = 10^{-2}$, $\lambda_{R} = 10^{-4}$ substantially improves the prototype geometry. We use two additional angular metrics to characterize the relation between prototype and embedding geometry. $RM^\circ$ is the root-mean-square aggregation of the angular distance to the closest neighboring prototype, while $P2M^\circ$ is the 2nd percentile of nearest-neighbor inter-class angles, reflecting the weakest 2\% of classes. For the jointly trained Fast-VGAN+ECAPA models analyzed in Section~IV-A, we additionally report $Intra^\circ$, defined as the mean angular distance between utterance-level embeddings belonging to the same speaker, and $NN^\circ$, defined as the mean angular distance from each classifier prototype to its nearest neighboring prototype. Comparing $Intra^\circ$ and $NN^\circ$ indicates whether the within-speaker angular spread of the embeddings remains smaller than the separation between neighboring classifier prototypes.

\subsection{Emergence of hyperspherical structure}
\label{sec:global}

\begin{table}[htbp]
\centering
\footnotesize
\setlength{\tabcolsep}{2.5pt}
\begin{tabular}{|l|c|c|c|c|c|c|}
\hline
\textbf{Model} &
\textbf{$\mathbf{RM^\circ \uparrow}$} &
\textbf{$\mathbf{P2M^\circ \uparrow}$} &
\textbf{$\mathbf{D_{PR} \uparrow}$} &
\textbf{$\mathbf{L_{PR} \uparrow}$} &
\textbf{$\mathbf{EER \downarrow}$} &
\textbf{$\mathbf{T5Acc \uparrow}$} \\
\hline
Init &
77.5 & 74.4 & 191.7 & 41.3 & -- & -- \\
\hline
ECA S1 &
20.142 & 7.941 & 15.2 & 4.7 & 0.040 & 0.622 \\
\hline
ECA S10 &
66.913 & 58.134 & 139.4 & 29.9 & 0.012 & 0.892 \\
\hline
ECA S20 &
72.355 & 66.699 & 165.6 & 35.0 & 0.099 & \textbf{0.924} \\
\hline
ECA S20 Reg &
\textbf{81.604} &
\textbf{80.544} &
\textbf{182.8} &
\textbf{40.0} &
0.109 &
0.890 \\
\hline
ECA S30 &
74.443 & 68.350 & 177.3 & 37.9 & 0.163 & 0.880 \\
\hline
\end{tabular}

\vspace{0.3cm}
\caption{Hyperspherical geometry and speaker-recognition performance for ECAPA-TDNN. ``S'' denotes the scale value and ``Reg'' the proposed regularization. The margin is 0.2 and the batch size is 32.}
\label{tab:hs_struc}
\vspace{-0.5cm}
\end{table}

As shown in \tabref{tab:hs_struc}, the classifier scale strongly influences the geometry of the hyperspherical prototype space. With a small scale ($S=1$), training produces a highly concentrated prototype distribution: both global and local participation ratios decrease drastically, while nearest-neighbor separation almost vanishes (RM$=20.1^\circ$, P2M$=7.9^\circ$). Increasing the scale progressively produces a better distributed prototype space. For $S=20$, the geometric metrics are substantially improved relative to smaller scales, together with strong classification performance. Regularization further increases angular separation and participation ratios, yielding the most isotropic prototype distribution. However, it slightly degrades EER and Top-5 accuracy relative to $S=20$. This illustrates that improved hyperspherical coverage does not necessarily translate into improved verification performance. Since EER primarily measures speaker discriminability rather than the suitability of the representation as a continuous conditioning space, we treat it here as a secondary diagnostic. The downstream VC experiments provide the more relevant evaluation for the main question of this work.
Interestingly, increasing the training batch size from 32 to 160 has only a marginal effect on the learned geometry. For the best-performing configuration ($S=20$), the geometric metrics remain very similar (RM: $72.4^\circ \rightarrow 74.3^\circ$, P2M: $66.7^\circ \rightarrow 69.2^\circ$, $G_{PR}$: $165.6 \rightarrow 164.7$, $L_{PR}$: $35.0 \rightarrow 36.9$), while the validation performance changes only slightly (T5Acc: $0.924 \rightarrow 0.911$, EER: $0.099 \rightarrow 0.076$). This weak dependence on batch size is particularly encouraging for applications such as Fast-VGAN \cite{abrassart2025fast}, where training is typically performed with a batch size of 32.

\section{Impact on Zero-Shot Voice-Conversion}

This section investigates the use of ECAPA-based speaker representations for zero-shot voice conversion and examines the relation between hyperspherical geometry and conversion performance. Audio samples and supplementary material are available online.\footnote{\url{https://abrassartm.github.io/Reg-VGAN/}}. We use Fast-VGAN \cite{abrassart2025fast} as the experimental backbone. We first vary the scale parameter of the ECAPA-TDNN angular-margin loss to analyze its effect on the geometry of the jointly trained speaker space. We then select a scale value and compare models trained with and without the proposed geometric regularization. Finally, we compare Fast-VGAN with four recent open-source zero-shot voice conversion baselines. The resulting systems are evaluated in terms of intelligibility, speaker similarity, perceptual quality, and generalization to unseen speakers.
Fast-VGAN is a voice conversion model based on a 2D convolutional generator that predicts a target Mel-spectrogram from temporally aligned conditioning features such as F0, energy, phoneme representations, and linguistic content. Further architectural and implementation details are provided in \cite{abrassart2025fast}.
The model is trained adversarially using a discriminator that distinguishes real from generated spectrograms. The resulting spectrograms are converted to waveforms using the neural vocoder MBExWN \cite{roebel2022neural}.
In the original any-to-many setting \cite{abrassart2025fast}, speaker identity is encoded via a learned lookup embedding. In contrast, zero-shot voice conversion requires embeddings extracted from a speaker encoder. We therefore use ECAPA-TDNN embeddings and study how the geometry of the corresponding classifier prototypes relates to generalization to unseen speakers.

\subsection{Hyperspherical structure}

Table~\ref{tab:hs_struc_fvg} reports the hyperspherical geometry metrics and speaker recognition performance for jointly trained Fast-VGAN and ECAPA-TDNN models under different scale values. For the best scale value, we added the result with regularization. 

\begin{table}[htbp]
\centering
\footnotesize
\setlength{\tabcolsep}{2.5pt}
\begin{tabular}{|l|c|c|c|c|c|c|c|}
\hline
\textbf{Model} &
\makecell{\textbf{$\mathbf{Intra^\circ}$}} &
\makecell{\textbf{$\mathbf{NN^\circ}$}} &
\makecell{\textbf{$\mathbf{P2M^\circ}$}\\$\mathbf{\uparrow}$} &
\makecell{\textbf{$\mathbf{D_{PR}}$}\\$\mathbf{\uparrow}$} &
\makecell{\textbf{$\mathbf{L_{PR}}$}\\$\mathbf{\uparrow}$} &
\makecell{\textbf{$\mathbf{EER}$}\\$\mathbf{\downarrow}$} &
\makecell{\textbf{$\mathbf{T5Acc}$}\\$\mathbf{\uparrow}$} \\
\hline
FVG-Init &
-- &
77.56 &
74.6 &
191.7 &
39.0 &
-- &
-- \\
\hline
FVG-S1 &
32.08 & 
16.57 &
8.20 &
8.9 &
9.0 &
0.079 &
0.50 \\
\hline
FVG-S15 &
56.01 &
63.73 &
59.1 &
50.0 &
24.5 &
0.031 &
0.87 \\
\hline
FVG-S20 &
59.19 &
65.95 &
62.9 &
54.8 &
27.1 &
0.029 &
0.89 \\
\hline
FVG-S30 &
61.37 &
68.19 & 
64.6 &
63.8 &
30.3 &
0.029 &
0.90 \\
\hline
FVG-S30 Reg &
66.52 &
81.69 &
\textbf{72.4} &
\textbf{118.7} &
\textbf{38.0} &
\textbf{0.026} &
\textbf{0.93} \\
\hline
FVG-S40 &
62.84 &
68.92 & 
61.35 &
61.75 &
28.7 &
0.029 &
0.89 \\
\hline
\end{tabular}

\vspace{0.3cm}
\caption{Hyperspherical geometry and speaker-recognition performance for Fast-VGAN+ECAPA (FVG). $Intra^\circ$ and $NN^\circ$ denote the mean intra-speaker and nearest-prototype angles, respectively; ``S'' is the scale value and ``Reg'' the proposed regularization.}
\label{tab:hs_struc_fvg}
\vspace{-0.5cm}
\end{table}

As observed with the standalone ECAPA-TDNN, modifying the scale parameter also affects the geometry of the learned speaker prototypes in the joint training setting. Increasing the scale from 15 to 30 progressively increases the geometric metrics, while the proposed regularization produces the highest participation ratios and angular separation among the evaluated configurations. In this joint Fast-VGAN setting, FVG-S30 Reg also yields the lowest EER (0.026) and highest Top-5 accuracy (0.93). This contrasts with the standalone ECAPA experiment, where regularization slightly degrades these recognition metrics, indicating that the effect on speaker-recognition performance depends on the training configuration.

Except for the strongly collapsed $S=1$ configuration, the mean intra-speaker angular distance remains smaller than the mean nearest-neighbor prototype angle. This indicates that, once sufficient inter-prototype separation is established, utterance-level embeddings form speaker-specific clusters whose angular extent remains smaller than the separation between neighboring classifier prototypes. The regularized $S=30$ model shows the largest margin between these quantities ($66.5^\circ$ vs. $81.7^\circ$), further supporting the correspondence between the geometry of the classifier prototypes and that of the utterance-level embeddings.

\subsection{Neighborhood organization with respect to speaker gender}

\begin{table}[htbp]
\begin{center}
\footnotesize
\setlength{\tabcolsep}{5pt}
\begin{tabular}{|l|c|c|c|c|c|c|}
\hline
\textbf{Model} & \multicolumn{6}{c|}{\textbf{Neighborhood size \(\mathbf{k}\)}} \\
\cline{2-7}
& \textbf{10} & \textbf{50} & \textbf{100} &
\textbf{200} & \textbf{300} & \textbf{500} \\
\hline
FVG S1 & $\bm{96.42}$ & $\bm{95.54}$ & $\bm{95.68}$ & $\bm{95.31}$ & $\bm{93.10}$ & $\bm{77.60}$ \\
\hline
FVG S15 & 94.28 & 85.49 & 75.94 & 64.45 & 57.98 & 51.09 \\
\hline
FVG S20 & 91.96 & 81.90 & 73.07 & 62.81 & 57.17 & 51.16 \\
\hline
FVG S30 & 89.45 & 78.62 & 70.71 & 61.55 & 56.78 & 51.44 \\
\hline
FVG S30 Reg & 81.83 & 74.18 & 69.44 & 64.68 & 61.47 & 57.12 \\
\hline
FVG S40 & 67.23 & 67.09 & 69.14 & 61.36 & 55.82 & 49.72 \\
\hline
\end{tabular}
\vspace{0.3cm}
\caption{Percentage of same-gender speakers among the $k$ nearest neighbors of each speaker prototype. Higher values indicate a stronger gender-based organization of the prototype space.}
\label{tab:gender_knn}
\end{center}
\vspace{-0.5cm}
\end{table}

As an additional analysis, Table~\ref{tab:gender_knn} reports the gender composition of local neighborhoods in the learned hyperspherical space. For each speaker prototype, we measure the fraction of same-gender speakers among its $k$ nearest neighbors.
Although gender is not used during training, the prototypes exhibit a strong gender-based organization (Table~\ref{tab:gender_knn}). This trend must be interpreted together with Table~\ref{tab:hs_struc_fvg}: $S=1$ collapses to $D_{PR}=8.9$ with very small inter-prototype separation, whereas S30 Reg reaches $D_{PR}=118.7$ and approaches the angular dispersion of random initialization. Thus, nearest neighbors do not represent comparable angular neighborhoods across configurations. Nevertheless, gender organization remains strong after regularization, with 81.8\% and 74.2\% same-gender speakers for $k=10$ and $k=50$, respectively.

\subsection{Experiments}
\label{section:multimedia}

\textbf{Datasets \& training}: Fast-VGAN is trained on the LibriTTS-R train-clean subset \cite{zen2019libritts} ($\sim$250 h, 1.1k speakers), with the augmentation recipe of \cite{desplanques2020ecapa} except reverberation. Zero-shot evaluation uses unseen VCTK speakers \cite{veaux2017cstr} at 22.5 kHz, with 20 gender-balanced conversion pairs (40 utterances) per condition. For EZ-VC \cite{joglekar2025ez}, Seed-VC \cite{liu2024zero}, GenVC \cite{cai2025genvc}, and kNN-VC \cite{baas2023voice}, we evaluate the authors' pretrained checkpoints on the same VCTK pairs using the same objective pipeline; VCTK is not among their reported training corpora.

\subsection{Objective VC evaluation}
\label{section:supplementary}

We evaluate intelligibility using WER from a pretrained NVIDIA NeMo model\footnote{\url{https://github.com/NVIDIA/NeMo/}}, speaker similarity using cosine similarity between Resemblyzer embeddings \footnote{\url{https://github.com/resemble-ai/Resemblyzer}}, and perceptual quality using UTMOSv2. Reported results include 95\% bootstrap confidence intervals.

\begin{table}[htbp]
\centering
\footnotesize
\setlength{\tabcolsep}{2.5pt}
\begin{tabular}{|l|c|c|c|}
\hline
\textbf{Model (Params)} &
\textbf{WER (\%) $\downarrow$} &
\textbf{Sim. $\uparrow$} &
\textbf{UTMOSv2 $\uparrow$} \\
\hline

Ground truth &
$2.33 \pm 2.43$ &
$0.77 \pm 0.03$ &
-- \\
\hline

Fast-VGAN S30 (9M) &
$6.54 \pm 3.26$ &
$0.65 \pm 0.03$ &
$2.47 \pm 0.17$ \\
\hline

Fast-VGAN S30 Reg (9M) &
$\bm{5.97 \pm 3.96}$ &
$0.69 \pm 0.02$ &
$2.70 \pm 0.19$ \\
\hline

EZ-VC \cite{joglekar2025ez} (910M) &
$16.93 \pm 4.32$ &
$\bm{0.75 \pm 0.03}$ &
$\bm{3.28 \pm 0.11}$ \\
\hline

Seed-VC \cite{liu2024zero} (440M) &
$\bm{4.84 \pm 1.87}$ &
$0.69 \pm 0.03$ &
$3.06 \pm 0.11$ \\
\hline

Gen-VC \cite{cai2025genvc} (620M) &
$10.17 \pm 3.50$ &
$0.66 \pm 0.03$ &
$2.38 \pm 0.12$ \\
\hline

KNN-VC \cite{baas2023voice} (330M) &
$16.71 \pm 3.88$ &
$\bm{0.73 \pm 0.03}$ &
$2.58 \pm 0.11$ \\
\hline
\end{tabular}

\vspace{0.3cm}
\caption{Objective VC evaluation. Lower WER is better, while higher similarity and UTMOSv2 scores indicate better performance.}
\label{tab:vc_results}
\vspace{-0.5cm}
\end{table}

Table~\ref{tab:vc_results} shows that regularization improves Fast-VGAN in WER (6.54\% to 5.97\%), similarity (0.65 to 0.69), and UTMOSv2 (2.47 to 2.70). Despite using only 9M parameters and $\sim$250 h of training data, Fast-VGAN remains competitive with much larger systems trained or pretrained on substantially larger corpora (12.8k h for EZ-VC, 101k h for Seed-VC, $>$46k h for GenVC-Large, and 94k h for the WavLM encoder used by kNN-VC).

\subsection{Subjective VC evaluation}
\label{section:repos}

We conducted a subjective evaluation to investigate whether the position of a target speaker in the speaker embedding space affects zero-shot voice conversion quality. A Mean Opinion Score (MOS) evaluation was conducted on the Prolific\footnote{\url{https://www.prolific.com/}} crowdsourcing platform with 50 participants. Listeners rated naturalness and speaker similarity on a 5-point Likert scale. Our hypothesis is that broader embedding-space coverage is associated with better generalization, particularly for speakers farther from the training distribution. VCTK speakers are ranked by their mean angular distance to the $k=35$ nearest training speakers and split into three equal-sized groups: \textit{Near}, \textit{Medium}, and \textit{Far}. Voice conversion samples were generated with both the baseline ECAPA system and the proposed ECAPAReg system. Table~\ref{tab:vctk_distance_mos} reports the results for the three distance groups, allowing us to assess whether embedding distance correlates with conversion difficulty and whether ECAPAReg provides larger improvements for speakers farther from the training distribution.

\begin{table}[htbp]
\centering
\footnotesize
\setlength{\tabcolsep}{3pt}
\begin{tabular}{|l|ccc|ccc|}
\hline
\multirow{2}{*}{\textbf{Model}} &
\multicolumn{3}{c|}{\textbf{Naturalness}} &
\multicolumn{3}{c|}{\textbf{Similarity}} \\
\cline{2-7}
 & \textbf{Near} & \textbf{Medium} & \textbf{Far}
 & \textbf{Near} & \textbf{Medium} & \textbf{Far} \\
\hline
FVG S30     
& $2.885$ & $2.682$ & $2.300$
& $2.545$ & $2.500$ & $1.962$ \\
\hline
FVG S30 Reg 
& $\bm{3.048}$ & $\bm{2.857}$ & $\bm{2.714}$
& $\bm{2.762}$ & $\bm{2.667}$ & $\bm{2.143}$ \\
\hline
\hline
Ground truth
& \multicolumn{3}{c|}{$3.961 \pm 1.231$}
& \multicolumn{3}{c|}{$4.351 \pm 1.106$} \\
\hline
\end{tabular}

\vspace{0.3cm}
\caption{Subjective MOS evaluation on VCTK speakers grouped according to their average angular distance to the $k$ nearest training speakers. Near, Medium, and Far denote increasing distance from the training speaker distribution.}
\label{tab:vctk_distance_mos}
\end{table}
\vspace{-0.3cm}

The results in Table~\ref{tab:vctk_distance_mos} show that voice conversion quality decreases as the target speaker becomes more distant from the training speaker distribution. The proposed regularized model consistently improves both naturalness and speaker similarity across all three distance groups. Moreover, the largest improvement in naturalness is observed for the \textit{Far} speakers, suggesting that the proposed regularization is particularly beneficial for speakers located farther from the training distribution. Overall, these results are consistent with the hypothesis that broader coverage of the speaker representation space is associated with better generalization to previously unseen speakers, although they do not establish a causal relation between isotropy and conversion quality.

\subsection{Interpolation across low- and high-density regions}

We next investigate regions of the speaker space that are not directly produced by the ECAPA-TDNN encoder for individual utterances, but are reached by interpolating between pairs of unseen VCTK speaker representations. For each speaker pair, we generate a SLERP trajectory between the two endpoint embeddings and evaluate the density of its interior with respect to the LibriTTS-R training-speaker prototypes. We define \textit{Void} trajectories as those whose intermediate points remain far from the nearest training prototype, whereas \textit{Dense} trajectories pass through regions closer to existing prototypes. The endpoints are excluded so that the density criterion characterizes the interpolated region itself rather than the observed VCTK speaker representations. To control for trajectory length, each Void trajectory is matched with a Dense trajectory having a similar angular separation between its two endpoint speakers.

\begin{figure}[h]
    \centering
    \includegraphics[width=\linewidth]{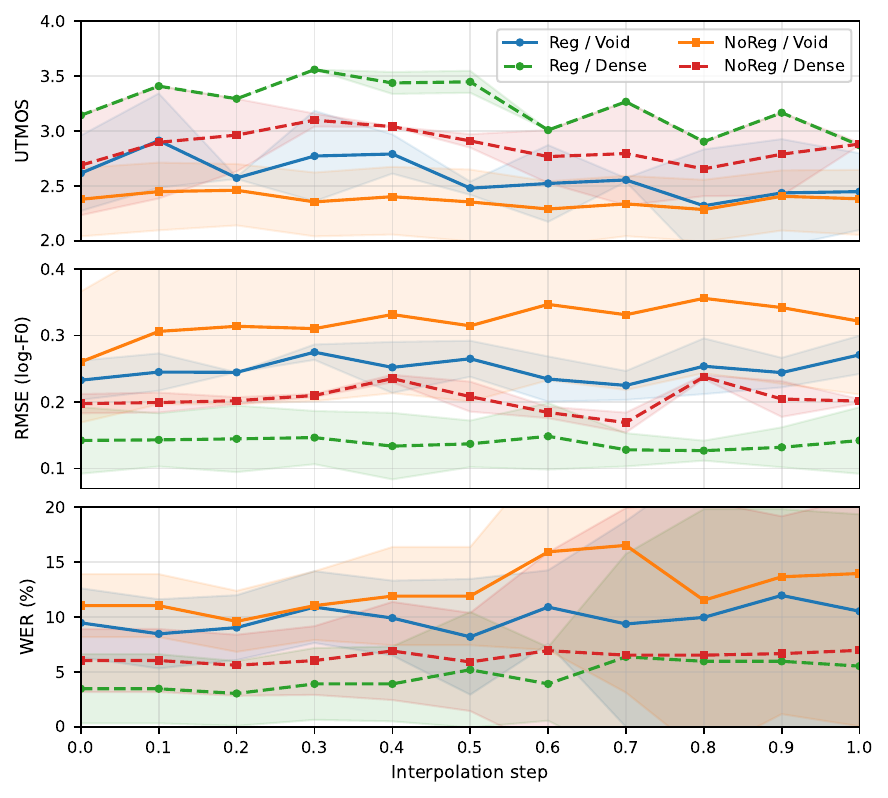}
    \caption{Interpolation performance in low-density (\textit{Void}) and high-density (\textit{Dense}) regions, with trajectory pairs matched for the angular separation between their endpoint speakers.}
    \label{fig:void_interp}
\end{figure}

As shown in Figure~\ref{fig:void_interp}, Dense trajectories generally yield higher UTMOS and lower log-F0 RMSE and WER. In Void regions, regularization improves UTMOS and log-F0 RMSE and tends to reduce WER, while it does not degrade the Dense controls. These results are consistent with improved robustness in poorly covered regions rather than a trade-off with well-covered regions.

\subsection{Interpolation from one-shot speakers}

We further investigate the robustness of the learned speaker space under limited enrollment using unseen speakers from VCTK. For each target speaker, the ECAPA-TDNN representation is extracted from a single short reference utterance of approximately one second. Interpolations are then generated between all pairs of these one-shot VCTK speaker representations using the same protocol as in the previous experiment.

\begin{figure}[h]
    \centering
    \includegraphics[width=\linewidth]{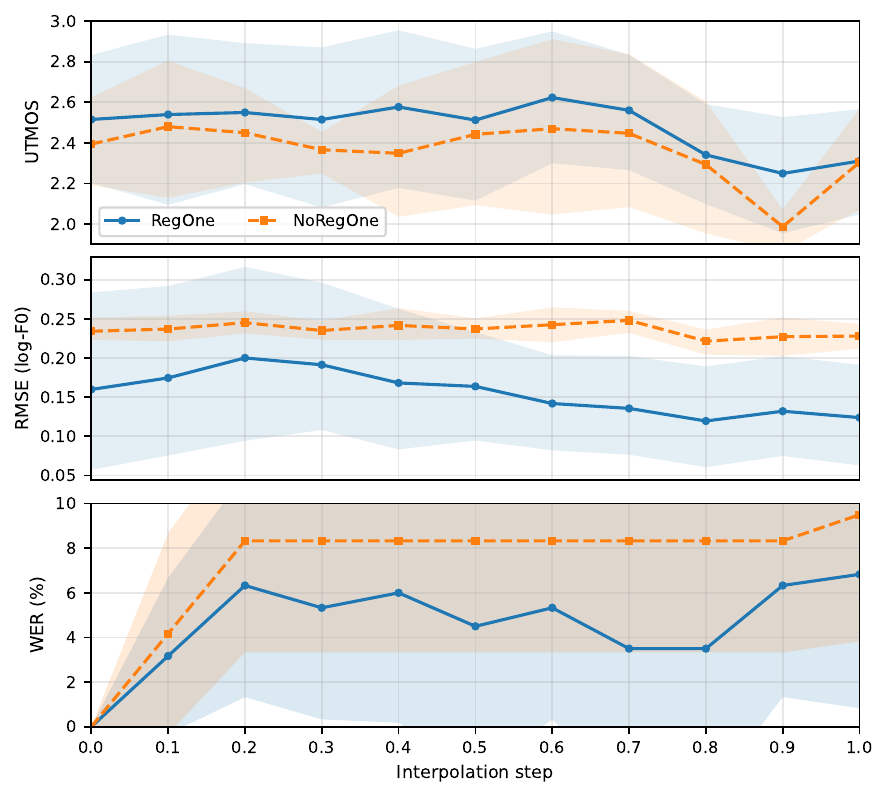}
    \caption{Interpolation quality for speaker pairs represented by a single short enrollment utterance.}
    \label{fig:oneshot_interp}
\end{figure}

Figure~\ref{fig:oneshot_interp} compares the baseline and the regularized models. The regularized model consistently achieves higher UTMOS scores and substantially lower log-F0 RMSE across the interpolation trajectory. It also produces lower WER for nearly all interpolation coefficients, indicating that speech intelligibility is better preserved despite the limited speaker information. These results associate a more uniform speaker space with more robust interpolation under limited enrollment.

\begin{figure}[h]
    \centering
    \includegraphics[width=\linewidth]{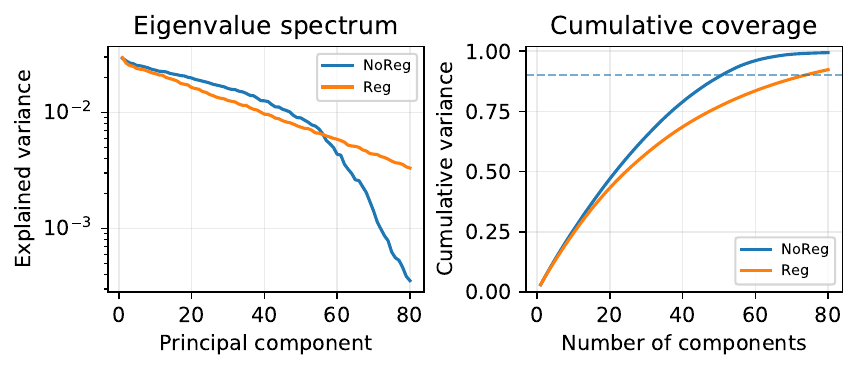}
    \caption{PCA of the prototype space. Regularization flattens the
    eigenvalue spectrum and increases $d_{90}$ from 45 to 65,
    indicating reduced anisotropy.}
    \label{fig:pca_geometry}
\end{figure}

PCA further confirms this effect: regularization flattens the
eigenvalue spectrum and increases the number of components required
to explain 90\% of the variance from 45 to 65.

\section{Discussion}

Our experiments show that angular-margin training can produce strongly structured prototype spaces with substantial variations in angular separability, anisotropy, and effective dimensionality. The proposed geometric regularizers promote a more isotropic and higher-dimensional hyperspherical distribution by increasing angular separation and distributing variance across a larger number of dimensions. These geometric changes are associated with improved robustness in zero-shot voice conversion, particularly for unseen speakers, while speaker-recognition effects depend on the training configuration. Overall, geometry appears to provide a useful prior for zero-shot conditioning; isolating the effect of the two regularizers and training-seed variability remains future work.

\section{Generative AI Use Disclosure}
Generative AI has been used for improving, correcting, and compacting the English language, as well as for scripting related to data analysis and the generation of plots.

\section*{Acknowledgment}

This work was partly performed using HPC resources from GENCI-IDRIS (Grant 2025-AD011011177R5), and partly funded by the ANR project BRUEL (ANR-22-CE39-0009).

\bibliographystyle{ieeetr}
\bibliography{mybib}

\end{document}